\documentclass[twocolumn,pra,showpacs,superscriptaddress]{revtex4}

\usepackage{mathrsfs}
\usepackage{graphicx}
\usepackage{dcolumn}
\usepackage{bm}
\usepackage{amsmath}
\usepackage{enumerate}
\usepackage{longtable}
\usepackage{arydshln}

\begin{document}

\title{Simulating dynamical quantum Hall effect with superconducting qubits}

\author{Xu-Chen Yang}
\affiliation{Laboratory of Quantum
Engineering and Quantum Materials, SPTE, South China Normal
University, Guangzhou 510006, China}\affiliation{National Laboratory of Solid
State Microstructures and School of Physics, Nanjing University,
Nanjing 210093, China}

\author{Dan-Wei Zhang}
 \email{zdanwei@126.com}
\affiliation{Laboratory of Quantum Engineering and Quantum
Materials, SPTE, South China Normal University, Guangzhou 510006,
China}\affiliation{Department of Physics and Center of Theoretical and Computational Physics, The University of Hong Kong,
Pokfulam Road, Hong Kong, China}

\author{Peng Xu}
\affiliation{National Laboratory of Solid
State Microstructures and School of Physics, Nanjing University,
Nanjing 210093, China}

\author{Yang Yu}
\affiliation{National Laboratory of Solid
State Microstructures and School of Physics, Nanjing University,
Nanjing 210093, China}

\author{Shi-Liang Zhu}
\email{slzhu@nju.edu.cn} \affiliation{National Laboratory of Solid
State Microstructures and School of Physics, Nanjing University,
Nanjing 210093, China}
%\affiliation{Laboratory of Quantum
%Engineering and Quantum Materials, SPTE, South China Normal
%University, Guangzhou 510006, China}

\begin{abstract}
We propose an experimental scheme to simulate the dynamical
quantum Hall effect and the related interaction-induced
topological transition with a superconducting-qubit array. We show
that a one-dimensional Heisenberg model with tunable parameters
can be realized in an array of superconducting qubits. The
quantized plateaus, which is a feature of the dynamical quantum Hall
effect, will emerge in the Berry curvature of the superconducting
qubits as a function of the coupling strength between nearest
neighbor qubits. We numerically calculate the Berry curvatures of
two-, four- and six-qubit arrays, and find that the
interaction-induced topological transition can be easily observed
with the simplest two-qubit array. Furthermore, we analyze some practical
conditions in typical experiments for observing such dynamical
quantum Hall effect.
\end{abstract}

\date{\today}

%\pacs{74.81.Fa, 03.65.Vf, 85.25.Cp, 03.67.Ac}
\maketitle

\section{introduction}

%\part{context}

The quantum Hall effect (QHE) is one of the most remarkable
phenomena in condensed matter physics \cite{Klitzing1980,Tsui}.
The basic experimental fact characterizing QHE is that the
non-diagonal conductivity is quantized in the form of $n e^2/h$
with $n$ being an integer (the integer QHE) or a fractional number
(the fractional QHE). The integer $n$ is a topological invariant
which can be expressed as the integral of the Berry curvature
\cite{Berry} over the momentum space \cite{Thouless1982,Niu1985}.
The Berry curvature and its associated Berry phase have many
additional applications in condensed matter physics
\cite{Xiao,Zhu2006} and quantum
computation \cite{Zanardi,Sjoqvist,Zhu2002}. Usually the Berry
phase is measured with the interference experiments. Recently, it
was proposed that the Berry curvature and hence the Berry phase in
generic systems can be detected as a non-adiabatic response on
physical observables to the rate of change of an external
parameter \cite{Gritsev2012,Avron2011}. This phenomenon can be
interpreted as a dynamical QHE in a parameter space, while the
conventional QHE is a particular example of the general relation
if one views the electric field as a rate of change of the vector
potential \cite{Gritsev2012}. This work opens up the possibility
to study the QHE in parameter space and to
measure the Berry phase in many-body systems.

On the other hand, superconducting qubits have become one of the
leading systems to study the Berry phase and to simulate some
interesting phenomena emerged in condensed matter
physics \cite{Georgescu2014}. The Berry phase \cite{Leek2007}, the
non-Abelian non-adiabatic geometric gates \cite{Abdumalikov}, and
the geometric Landau-Zener interference \cite{Tan} were
experimentally demonstrated with superconducting qubits.
Furthermore, a topological transition characterized by the change
of the Chern number was also experimentally observed
\cite{Schroer2014,Roushan2014}. These studies suggest that the
superconducting qubit system can be a promising system for further
exploring rich topological features of single-particle and many-body
physics.

In this paper, we propose an experimental scheme to simulate the
dynamical QHE and the related interaction-induced topological
transition with a superconducting-qubit array. The one-dimensional (1D)
Heisenberg spin chain was proposed to realize with superconducting
qubits \cite{Pinto2010,Paraoanu2014}. We first extend this
approach to show that an almost isotropic interaction (i.e.,
$J^x_j=J^y_j \approx J^z_j$ in Eq. (4)) between the nearest
neighbor superconducting qubits can be achieved by coupling phase qubits with the Josephson
junctions controlled with the bias current. One of the advantages
of the system is that all parameters in this 1D
Heisenberg model are controllable and tunable in experiments. We
then show that the dynamical QHE and the related
interaction-induced topological transition can be observed in the
system. We numerically calculate the Berry curvatures of two-, four-
and six-qubit arrays, and find that the interaction-induced
topological transition can be easily observed with the simplest
two-qubit array. We also discuss some practical conditions for observing the dynamical QHE in this system, such
as the limit of ramp velocity, the control errors in
experiments and the decoherence effects for the realistic open-system conditions.

The rest of this paper is organized as follows:
Section II introduces our proposed superconducting phase qubit
array and the realization of the required spin Hamiltonian. Section III presents our results for
observing the dynamical QHE and the related interaction-induced
topological transition in the proposed system. In Sec. IV, we analyze the
ramp velocity limit, the robustness of our scheme
against the control errors and the decoherence effects for realistic conditions, and finally present our conclusions.

\begin{figure}[tbp]
\includegraphics[width=7.0cm]{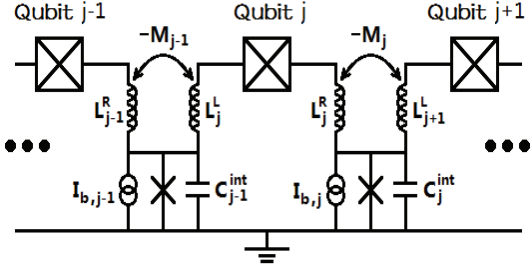}
\caption{Schematic diagram of the superconducting qubits array to simulate
the dynamical QHE. The nearest-neighbor phase
qubits, such as qubit $j$ and qubit $j+1$, are
coupled through a Josephson junction (denoted by $\times$ in the figure) with capacitance $C_j^{\rm int}$. The two qubits also contain the capacitances $C_j$ and $C_{j+1}$, while the circuit has a negative mutual
inductance $-{M_j}$ for the inductors $L_j^{R}$ and $L_{j + 1}^{L}$ and a tunable bias current ${I_{b,j}}$. }
\end{figure}

\section{system and Hamiltonian}

It was demonstrated that the dynamical QHE can emerge in a $1$D
Heisenberg spin chain model with tunable parameters \cite{Gritsev2012}.
We consider the $1$D Heisenberg spin chain
model with an external magnetic field:
\begin{equation}
\label{H_H}
H =  - \sum\limits_{j = 1}^N {\vec h \cdot {{\vec
\sigma }_j}}  + J\sum\limits_{j = 1}^{N - 1} {{{\vec \sigma
}_j} \cdot {{\vec \sigma }_{j +
  1}}},
\end{equation}
where $\vec \sigma\equiv(\sigma_x,\sigma_y,\sigma_z)$ stands for Pauli matrices,
$J$ is the isotropic coupling constant between the nearest-neighbor spins,
$\vec h\equiv(h^x,h^y,h^z)$ is the external magnetic field, and $N$ is the size of the spin chain.
In the following, we will show that this Hamiltonian with tunable coupling constants can be realized in an array of
superconducting phase qubits and the related dynamical QHE can be
observed in this system.

The schematic diagram of the whole system we consider is shown in
Fig 1. It is an array of $N$ superconducting phase qubits coupled
with Josephson junctions. The phase qubit $j$ constitutes an
"atom-like" two-level system. The truncated Hamiltonian of the
lowest two levels $(|0\rangle,|1\rangle)$ in the energy bases is
$H_q=\frac{1}{2}\hbar \omega_q \sigma_z$, where $\hbar\omega_q$
represents the energy difference between $|0\rangle$ and
$|1\rangle$ and $\sigma_z$ is the Pauli operator in the $z$
direction \cite{Orlando1999,Tan}. For simplicity, we assume the same parameters for all the phase qubits (i.e., ${\omega
_{q,j}} ={\omega _q}$). Moreover, the state of each qubit can be
controlled by microwaves. In the rotating frame of an applied
microwave with the frequency ${\omega _d}$, the Hamiltonian for
the qubits can be written as \cite{Georgescu2014,Paraoanu2014}
\begin{equation}
  H =  - \sum\limits_{j = 1}^N {\vec h \cdot {{\vec \sigma }_j}}  + {H_{{\mathop{\rm int}}
  }}.
\end{equation}
Here the interacting part of the Hamiltonian $H_{{\mathop{\rm
int}}}$ will be addressed later, and $\vec h $ is an effective magnetic
field induced by the microwave and can be parameterized as
\cite{Schroer2014,Roushan2014,Tan}
\begin{equation}
\begin{split}
h^x\left( t \right) &= {h}\sin \theta  \cos  \phi  ,\\
h^y\left( t \right) &= {h}\sin  \theta \sin  \phi  ,\\
h^z\left( t \right) &= {h}\cos \theta.
\end{split}
\end{equation}
Here the parameter $\phi$ represents the
phase of the applied microwave, $h\sin\theta$ is the Rabi oscillation frequency
proportional to the amplitude of the microwave, and $h \cos\theta=\omega _d-\omega _q$
is the detuning with $\omega_d$ being the frequency of the microwave and $\theta$ being the mixing angle. The
mixing angle will be used as the quench parameter
for observing the dynamical QHE in the following.

As shown in Fig. 1, the interaction between nearest-neighbor qubits $j$ and
$j+1$ is realized by the inductances $L_j^{R}$ and $L_{j +
1}^{L}$ and the Josephson junction characterized by
capacitance $C_j^{{\mathop{\rm int}} }$. The two qubits also contain the capacitances $C_j$ and $C_{j+1}$, while the circuit has a negative mutual
inductance $-{M_j}$ and a tunable
bias current ${I_{b,j}}$. Thus in this system, the coupling strengths can be tuned
via the bias current of the coupled
Josephson junctions ${I_{b,j}}$ and the Hamiltonian of the interacting part
${H_{{\mathop{\rm int}} }}$ can be written as \cite{Pinto2010,Paraoanu2014}
\begin{equation}
  {H_{{\mathop{\rm int}} }} = \sum\limits_{j = 1}^{N - 1} {\left( {J_j^x\sigma _j^x\sigma _{j + 1}^x +
  J_j^y\sigma _j^y\sigma _{j + 1}^y + J_j^z\sigma _j^z\sigma _{j + 1}^z}
  \right)},
\end{equation}
where the coupling strengths along the three spin directions are respectively given by \cite{Pinto2010}
\begin{align}
  J_j^x &= J_j^y = \frac{{{{\tilde M}_j} - {{\tilde L_j^{{\mathop{\rm int}} }} \mathord{\left/
 {\vphantom {{\tilde L_j^{{\mathop{\rm int}} }} {\left[ {1 - {{\left( {{{{\omega _q}} \mathord{\left/
 {\vphantom {{{\omega _q}} {\omega _j^{{\mathop{\rm int}} }}}} \right.
 \kern-\nulldelimiterspace} {\omega _j^{{\mathop{\rm int}} }}}} \right)}^2}} \right]}}} \right.
 \kern-\nulldelimiterspace} {\left[ {1 - {{\left( {{{{\omega _q}} \mathord{\left/
 {\vphantom {{{\omega _q}} {\omega _j^{{\mathop{\rm int}} }}}} \right.
 \kern-\nulldelimiterspace} {\omega _j^{{\mathop{\rm int}} }}}} \right)}^2}} \right]}}}}{{\tilde L_j^{R}\tilde L_{j + 1}^{L}{\omega _q}\sqrt {{C_j}{C_{j + 1}}} }},\\
  J_j^z &= \frac{1}{{6\sqrt {{N_{1,j}}{N_{2,j}}} }}\frac{{{{\tilde M}_j} - \tilde L_j^{{\mathop{\rm int}} }}}{{\tilde L_j^{R}\tilde L_{j + 1}^{L}\omega_q \sqrt {{C_j}{C_{j + 1}}} }}.
\end{align}
Here $\tilde L_j^{\rm int} = L_j^{\rm int}\left( {1 +
\frac{{{M_j}}}{{L_j^{R}}}} \right)\left( {1 + \frac{{{M_j}}}{{L_{j
+ 1}^{L}}}} \right)$ and $\omega _j^{{\rm int} } = 1/\sqrt{L_j^{{\rm int}}C_j^{{\rm int}}}$ with $L_j^{\rm int}=1/\sqrt{I_{j,cr}^2 - I_{b,j}^2}$ and $I_{j,cr}$ being the critical current of
inter-Josephson junction, the renormalization parameters (the mutual and coupling
inductances) are
$\frac{{{{\tilde M}_j}}}{{{M_j}}} = \frac{{\tilde L_j^{R}}}{{L_j^{R}}} = \frac{{\tilde L_{j + 1}^{L}}}{{L_{j + 1}^{L}}} =
 1 - \frac{{M_j^2}}{{L_j^{R}L_{j + 1}^{L}}}$, and $N_{1,j}$ ($N_{2,j}$) is
crudely the number of energy levels in the energy potential well for qubit $j$ ($j+1$),
which is about 5 in typical experiments with phase qubits.

From Eqs. (5,6), we calculate the coupling coefficients $J_j^x$ and
$J_j^z$ for the typical homogenous
parameters $\omega_q=4.77$ GHz, ${C_j} = 1.0 $ pF, ${L_j}
= 0.7 $ nH, $L_j^{L} = L_j^{R} = 3.0 $ nH,
${M_j} = 0.41 $ nH, ${N_{1,j}} = {N_{2,j}} = 5$ and ${I_{j,cr}} = 3.0~\mu $A \cite{Pinto2010}.
In this case, we get homogenous (qubit-independent) coupling strengths and thus the
qubit label $j$ in the coupling strengths $J^{r}_{j}$ ($r=x,y,z$) will be omitted for simplicity hereafter. The results are plotted in Fig. 2,
showing that if we adjust the bias current from 0 to $0.93{I_{j,cr}}$, the coupling strength $J^x$
will continuously and monotonically decrease from about $40$ MHz to zero. Furthermore, when the bias current is less than the
critical current $0.54{I_{j,cr}}$ the ratio $J^z/J^x$ remains in the region about $\left[ {0.9,1} \right]$.
As we will show in the following, this parameter region already allows the observation of the dynamical QHE.

\begin{figure}[tbph]
\includegraphics[width=8cm]{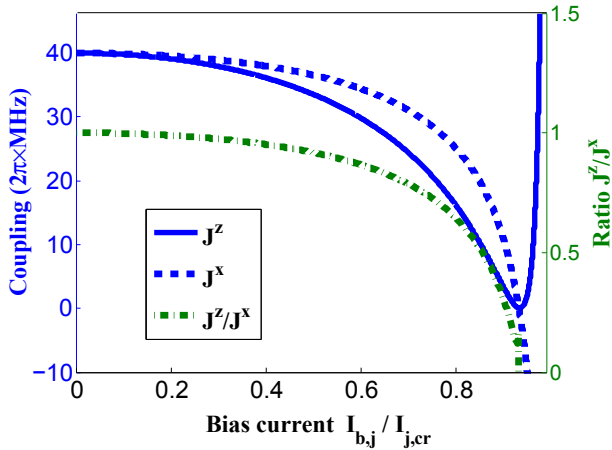}
  \caption{(Color online) The strength of coupling $J^x$ (blue dashed line), $J^z$ (blue solid line)
  and the ratio $J^z/J^x$ (green dashed-dotted line) as functions of the bias current $I_{b,j}/I_{j,cr}$.
  The typical parameters are $\omega_q=4.77$ GHz, ${C_j} = 1.0 $pF, ${L_j} = 0.7 $nH, $L_j^{L} = L_j^{R} = 3.0 $nH,
  ${M_j} = 0.41 $nH, ${N_{1,j}} = {N_{2,j}} = 5$, and ${I_{j,cr}} = 3.0 \mu $A.}
\end{figure}

\section{Simulating dynamical quantum Hall effect}

The topological features of the superconducting qubit system can
be probed by measuring the Berry curvature, while Fig. 3 depicts a
typical sequence used to measure the Berry curvature. To
demonstrate the dynamical QHE in this system, we follow the
proposal in Ref. \cite{Gritsev2012} to analyze the quantized
response of the system to a rotating magnetic field. We consider
all the superconducting qubits initially in the ground state with
$\theta(t=0)=0$, and then ramp the system with fixed $\phi(t)=0$
to undergo a quasi-adiabatic evolution by varying the mixing angle
$\theta(t)=v^2t^2/2\pi$ for a ramp time $t_{\text{ramp}}=\pi/v$,
where $v$ denotes the ramp velocity. At the end of such a ramp,
the velocity of the $\theta$-component of the magnetic field
$v_\theta(t)$ is exactly $v$, and we can measure the Berry
curvature of the system. We note that this choice of ramping field
guarantees that the angular velocity is turned on smoothly and the
system is not excited at the beginning of the evolution \cite{Gritsev2012}.

During the ramping process, the three components of
the effective magnetic field $h^{x,y,z}$ are depicted in Fig. 3 (a). The generalized
force for the full Hamiltonian $H$, which is measured at $t=t_{\text{ramp}}$, is along the latitude direction (at the point of measurement it is along
$y$-axis) and given by
 \begin{equation}
 M_\theta = -\langle \partial_{\phi} H \rangle \mid_{\phi=0,t=\pi/v} = h\sum_{j=1}^{N}\langle\sigma_j^y\rangle,
 \end{equation}
while the quench velocity is along to the longitude direction. Then we can
obtain the Berry curvature $F_{\theta \phi}$ within the linear response approximation \cite{Gritsev2012,Avron2011},
 \begin{equation}
   {F_{\theta \phi }} = \frac{M_\theta}{{{v_\theta }}} = \frac{h}{v}\sum\limits_{j = 1}^N {\left\langle {\sigma _j^y} \right\rangle }.
 \end{equation}
In experiments, one can measure the $\left\langle {\sigma _j^y}
\right\rangle$ of each superconducting qubit, and then the Berry
curvature can be derived by substituting the results into Eq. (8).
In other words, the qubits system in the described evolution
progress is initially prepared in the ground state and then is
slowly and smoothly driven along the $\theta$ direction, as shown
in Fig. 3(b). The generalized force $M_\theta$ along the
orthogonal direction is measured as a linear response to the
ramping magnetic field. We will see the quantization of this
response in the following, and in this sense it is called
dynamical QHE \cite{Gritsev2012}.

\begin{figure}[tbph]
  \centering
\includegraphics[width=8cm]{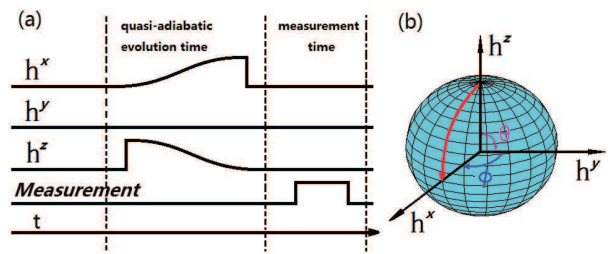}
  \caption{(Color online) (a) The schematic sequence diagram of the parameter evolution. The superconducting qubits undergo a non-adiabatic evolution, with the amplitude of effective magnetic field strength followed by Eq. (3). Then the $\left\langle {\sigma _j^y} \right\rangle$ of the final state is measured. (b) The schematic diagram of effective magnetic field strength. The red curve represents the evolution path. }
\end{figure}

\begin{widetext}

\begin{figure}[tbph]
\begin{center}
 \includegraphics[width=12cm]{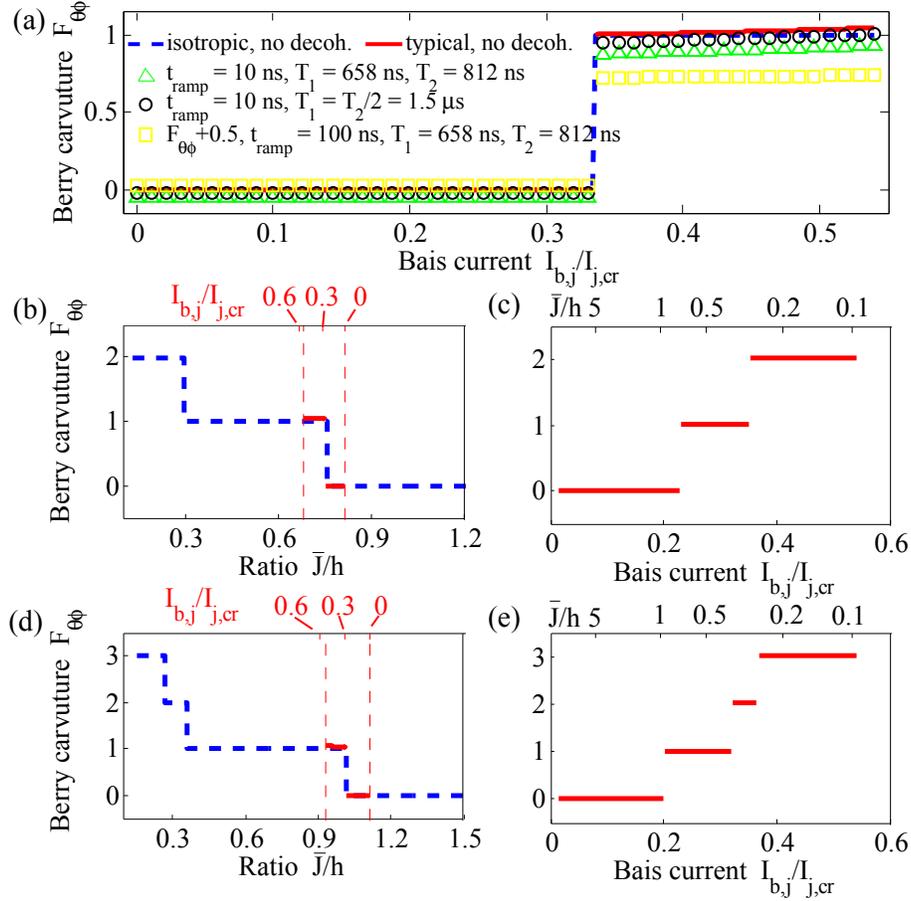}
  \caption{(Color online) The Berry curvature as a function of the ratio $\bar{J}/h$ (i.e., the ratio between the coupling strength and the amplitude of the effective magnetic field) or the bias current $I_{b,j}/I_{j,cr}$ in a two-, four- or six-qubit array. (a) Two-qubit case for $h/2\pi =76$ MHz. The red solid line is the Berry curvature as a function of $I_{b,j}/I_{j,cr}$, and the blue dashed line is the result for isotropic coupling strength
  $\bar{J}= \sqrt {(J^x)^2 + (J^y)^2 + (J^z)^2}/\sqrt{3}$ (here $J^{x,y,z}$ are determined by $I_{b,j}/I_{j,cr}$ as shown in Fig. 2). The green triangles and yellow squares are the corresponding $F_{\theta\phi}(I_{b,j}/I_{j,cr})$ in the open-system conditions for the decoherence times $T_1=658$ ns and $T_2=812$ ns, while the black circles for the case with $T_1=T_2/2=1.5$ $\mu$s, with the total measurement time $t_{\text{meas}}=10$ ns. (b) Four-qubit case for $h/2\pi =49$ MHz. (c) Four-qubit case for varied $h=-85\bar{J}+3400$ MHz. (d) Six-qubit case for $h/2\pi =36$ MHz. (e) Six-qubit
  case for varied $h=-85\bar{J}+3400$ MHz. In (b-e), the red solid and blue dashed lines represent the Berry curvature
  as a function of $I_{b,j}/I_{j,cr}$ and $\bar{J}/h$, respectively. The ramp time in (a-e) is $t_{\text{ramp}}=100$ ns except that $t_{\text{ramp}}=10$ ns for the green triangles and black circles in (a). We note that the results in (a-e) for the cases without decoherece almost remain for varying $t_{\text{ramp}}$ when $t_{\text{ramp}}\gtrsim 10$ ns.
  }\label{Fig4}
  \end{center}
\end{figure}

\end{widetext}

The simplest system to observe the dynamical QHE and its related
interaction-induced topological transition should be a two-qubit
system. Therefore we first address whether one can observe this
phenomenon in an array with two superconducting qubits. To this
end, we numerically calculate the Berry curvature $F_{\theta\phi}$
in a two-qubit array as a function of the bias current
$I_{b,j}/I_{j,cr}$ for the described ramp process by time-dependent exact diagonalization \cite{ED}.
The results are plotted in Fig. 4(a), where the parameters are the same with those in Fig. 2. The red solid line
in Fig. 4(a) shows that although the interaction strengths $J^{x,y,z}$ are not exactly isotropic in our superconducting qubit system,
the plateaus in the Berry curvature are strictly stable at $0$ and $1$,
and the transition between the two plateaus is very sharp. For
comparison, we also calculate the Berry curvature for the isotropic coupling case [the blue dashed line in Fig. 4(a)],
where we choose an isotropic coupling strength
$\bar{J}$ determined by $\bar{J}
= \sqrt {(J^x)^2 + (J^y)^2 + (J^z)^2}/\sqrt{3}$ (here $J^{x,y,z}$
depend on the bias current $I_{b,j}/I_{j,cr}$ as shown in Fig. 2). From Fig. 4(a), it is
clear that for the chosen typical parameters the difference of the Berry curvatures
between the isotropic and anisotropic cases can be neglected.

\begin{figure}[tbph]
 \includegraphics[height=4.5cm]{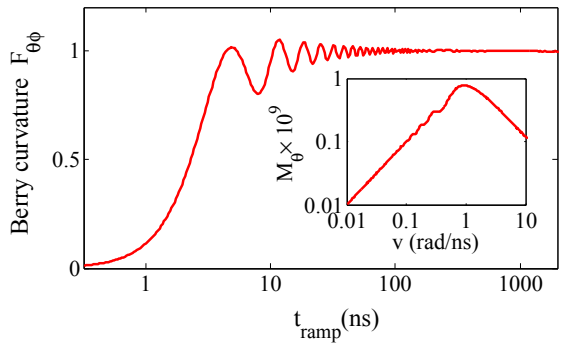}
  \caption{(Color online) The Berry curvature $F_{\theta\phi}$ as a function of the ramp time for a two-qubit array
  with isotropic coupling $\bar{J}=0.4h$ and ${h}/2\pi = 76$ MHz. The Berry curvature saturates to nearly
   one when $t_{\text{ramp}}\gtrsim 10$ ns. The inset shows the magnetization $M_\theta$ as a function of the finial ramp velocity $v$.
  }\label{Fig.5}
\end{figure}

Now we turn to address the dynamical QHE and the
interaction-induced topological transition in an $N$-qubit array.
For this 1D Heisenberg spin chain, the
plateaus in the Berry curvature should appear in an integer
$n=0,1,2...,N/2$ for an even $N$ \cite{Gritsev2012}. We have
numerically confirmed this phenomenon for a four- and six-qubit array,
with $F_{\theta\phi}(\bar{J}/h)$ for typical parameters shown as the blue dashed lines in Figs. 4(b) and 4(d),
respectively. We then further check whether these multi-plateaus can be observed in this superconducting
qubit system. For the same corresponding parameters, we calculate the Berry curvature as a
function of $I_{b,j}/I_{j,cr}$ in the region $I_{b,j}/I_{j,cr}\in[0,~0.54]$,
which corresponds to the coupling strength in the region [34,~40] MHz as shown in Fig. 2.
The results are given by the red lines in Figs. 4(b) and 4(d), where we find that only two
quantized plateaus with a topological transition appear and other quantized
plateaus could not be observed. One simple
approach to solve this problem is to simultaneously change the
magnetic field strength $h$ and the bias current ${I_{b,j}}$. For
instance, we can choose $h(\bar{J})=-85 \bar{J} + 3400$ MHz in simulations, and then the obtained Berry
curvature as a function of $\bar{J}/h$ (and $I_{b,j}/I_{j,cr}$) for
four-qubit and six-qubit are shown in Figs. 4(c) and 4(e),
respectively. It is clear that all
quantized plateaus can be observed in this approach. In the above calculation, the
amplitude of $h$ given by the relation equation $h(\bar{J})$ is yet to be optimized and it is about $350$ MHz at
$\bar{J}/h=0.1$. However, the required amplitude of the effective magnetic field
to observe all the quantized plateaus can be much smaller after the
optimization.

\section{Discussions and conclusions}

%%%%%%%%%%%%%%%%%%%%%%%%%%%%%%%%%%%%%%%%%%%%%%%%%%%
\begin{figure}[tbph]
 \includegraphics[height=3.9cm]{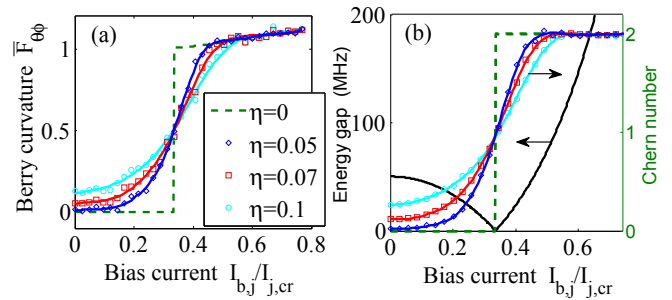}
  \caption{(Color online)  (a) The averaged Berry curvature $\bar{F}_{\theta\phi}$ of two qubits as a function of  ${I_{b,j}}/I_{j,cr}$ in the presence of
  fluctuating parameters $\tilde{J}^{x,y,z}$ and $\tilde{h}$. The fluctuation strengths are $\eta=0$ (green dashed line), $5\%$ (blue diamand), $7\%$ (red square) and $10\%$ (cyan circle), respectively. The other parameters are $h/2\pi =76$ MHz and $N_\alpha=500$. The ramp time is $t_{\text{ramp}}=100$ ns. (b) The average Chern number $Ch$ of two qubits with the same parameters in (a). In addition, the black line denotes the energy gap between the ground state and the first excited state of the system as a function of ${I_{b,j}}/I_{j,cr}$.
  }\label{Fig.6}
\end{figure}
%%%%%%%%%%%%%%%%%%%%%%%%%%%%%%%%%%%%%%%%%%%%%%%%%%%%%%%

In the previous calculations, the Berry curvature $F_{\theta\phi}$
is considered to be a linear response to the ramp velocity
$v_{\theta}$. In general, the magnetization (the generalized
force) is determined by $M_\theta=M_0+F_{\theta\phi}
v_\theta+\mathcal{O}(v_\theta^2)$ \cite{Gritsev2012,Avron2011},
where the constant term $M_0$ gives the value of the magnetization
in the adiabatic limit and $M_0=0$ in our cases. The linear
response theory breaks down when the velocity $v_{\theta}$ is too
large to neglect the term related to $v_\theta^2$. To check the
velocity limit in this linear response theory, we numerically
calculate the Berry curvature $F_{\theta\phi}$ as a function of
the ramp time $t_{\text{ramp}}$ for a two-qubit array, with the
results for parameters $\bar{J}/h=0.4$ and $h/2\pi = 76$ MHz being
plotted in Fig. 5. We can see that the Berry curvature saturates
to nearly one when $t_{\text{ramp}}\gtrsim 10$ ns and becomes very
stable when $t_{\text{ramp}}\gtrsim 60$ ns. In addition, the
magnetization $M_\theta$ is plotted in the inset of Fig. 5 as a
function of the finial ramp velocity $v$, which further shows the
linear response approximation works well within $v\lesssim
0.3~{\rm rad}/ \mu {\rm s}$. Therefore, to observe the quantized
plateaus in Fig. 4(a), the ramp velocity should be slower than
$0.3~{\rm rad}/ \mu {\rm s}$, corresponding to the ramp time
longer than $10$ ns. We also simulate the same procedures for the
four-qubit and six-qubit arrays and find that the results are
similar to those in Fig. 5. Thus the velocity limit to observe the
quantized plateaus does not change much for arrays with different
number of qubits.

Then we further study the robustness of the quantized plateaus of the Berry
curvature against the control errors which stem from the fluctuations
of the parameters in the Hamiltonian (2). We assume
$\tilde{J}^{x,y,x}=\alpha_1 J^{x,y,z}$ and $\tilde{h}=\alpha_2 h$, with
$\alpha_1$ and $\alpha_2$ randomly distributing in the region
$[1-\eta,1+\eta]$ (here $\eta>0$ describes the fluctuation strength). For a single realization with randomly chosen $\alpha_{1}$ and $\alpha_{2}$, we calculate
the corresponding $F_{\theta\phi}^\alpha$ as that in Fig. 4(a) and
then we can obtain the averaged Berry curvature $\bar{F}_{\theta\phi}=1/N_\alpha\sum
F_{\theta\phi}^\alpha$, where $N_\alpha$ denotes the sampling number. The averaged Berry curvature
for the two-qubit case as a function of ${I_{b,j}}/I_{j,cr}$ is
plotted in Fig. 6(a). We find that the plateaus
are still stable when the parameter fluctuation strength $\eta$ is less
than about $5\%$, even though their transition is slightly smoothed by the
fluctuation. Furthermore, we calculate the corresponding Chern number $Ch=(2\pi)^{-1}\int^{\pi}_0d\theta\int^{2\pi}_0d\phi F_{\theta\phi}=\int^{\pi}_0F_{\theta\phi}d\theta$ by integrating the Berry curvature in the $\theta-\phi$ sphere in Fig. 6(b).
As we expected, the Chern number is more robust and the quantized plateaus there are more significant due to the averaging over different runs of $\theta$-ramping with the parameter fluctuations. In addition, we also plot the energy gap between the ground state and the first excited state of the two-qubit array in Fig. 6(b). We can see that the gap closes at the topological transition point.

We now turn to discuss the decoherence effects in our
system for the realistic open-system conditions. For simplicity,
we assume that each superconducting qubit of the system interacts
independently with the environment, which is commonly modeled as a
bath of oscillators. The quantum dynamics of the system is thus
described by the master equation \cite{book}
\begin{equation}
\label{ME}
\frac{d\rho}{dt}=-\frac{i}{\hbar}[H,\rho]+\sum_{j=1}^{N}\mathcal{L}_j[\rho],
\end{equation}
where the density matrix $\rho$ is spanned by the $N$-qubit basis and the
Lindblad superoperator $\mathcal{L}_j[\rho]$ describes the decoherence due to the independent interaction between each qubit and the bath.
We further assume the weak qubit-bath interaction and the Markovian limit and thus the Lindblad superoperator can be
written as
$\mathcal{L}_j[\rho]=\gamma(1+n_0)(2\sigma_j^{-}\rho\sigma_j^{+}-\{\sigma_j^{+}\sigma_j^{-},\rho\})+\gamma n_0(2\sigma_j^{+}\rho\sigma_j^{-}-\{\sigma_j^{-}\sigma_j^{+},\rho\})+\Gamma(2\sigma_j^{z}\rho\sigma_j^{z}-\{\sigma_j^{z}\sigma_j^{z},\rho\})$ \cite{book}.
Here the first two terms describe the energy relaxation progress with parameter $\gamma$ and the third term describe the pure dephasing progress with parameter
$\Gamma$, and the effective boson number $n_0$ on each qubit depends on the
temperature of the bath $T$ with $n_0=1/[\exp(\hbar\omega_q/k_BT)-1]$ (here $k_B$ is the Boltzmann constant). In
superconducting qubit system, we have $n_0\approx0$ because $\hbar\omega_q\gg k_BT$ for $T\approx30$ mK and $\omega_q$ is on the order of gigahertz in practical experiments \cite{Tan,Schroer2014,Roushan2014}. Then the usually measured relaxation time $T_1$ and dephasing time $T_2$ of each qubit are determined by $1/T_1=\gamma$ and $1/T_2=1/2T_1+\Gamma$ \cite{book}, respectively.

The additional timescale for measurement is another issue one
should consider for finite decoherence time. We assume that each
phase qubit in the array can be manipulated and measured
independently \cite{Roushan2014,Lucero2012}. Since the qubits can
only be naturally read out in the $\sigma_z$ basis (i.e. the
$\langle\sigma^z\rangle$ measurement), an additional spin rotation
$$\hat{R}=\frac{1}{\sqrt{2}}\left(
                                                      \begin{array}{cc}
                                                        1 & -i \\
                                                        -i & 1 \\
                                                      \end{array}
                                                    \right)
$$ for each qubit (effectively an $\hat{X}_{\pi/2}$ operation in experiments \cite{Schroer2014,Roushan2014})
have to be inserted in order to measure $\langle\sigma^y_j\rangle$
after the ramp. This rotation can be achieved by additional
microwave pulses \cite{Tan,Roushan2014} with the duration time
$\tau_R\approx\pi/2h=3.3$ ns for the cases with $h/2\pi=76$ MHz in
Figs. 4 and 5. Finally, the $\langle\sigma^z_j\rangle$ measurement
of each qubit requires a duration time $\tau_d$, which is
typically several nanoseconds \cite{Tan,Lucero2012}. So the total
time required for measurement is around
$t_{\text{meas}}=\tau_R+\tau_d\approx10$ ns. Since the measurement
fidelity for each phase qubit in the coupled system can be more
than $95\%$ \cite {Lucero2012}, we do not further consider the
measurement errors.

To see the decoherence effects in the dynamical QHE in our
proposed system, we take the two-qubit array for example and
numerically simulate the whole progress with the ramp and
measurement sequences by calculate the master equation (see the
Appendix for details). For simplicity in our simulations, we treat
the evolution of the qubits in the whole measurement progress with
time $t_{\text{meas}}=10$ ns as free evolution under decoherence. In
addition, the relaxation time $T_1$ and dephasing time $T_2$ of
single phase qubit are usually longer than those of multi-qubit in
the coupled system; this effect is somehow contained in the master
equation, which includes the increases of decoherence channels and
decoherence rates (see Eq. (A8) in the Appendix). We estimate that
the effective times $\widetilde{T}_1$ and $\widetilde{T}_2$ of two qubits are about $5$
times smaller than those of a single qubit in the master equation (${T}_1$ and ${T}_2$),
thus we will choose typical decoherence times in simulations from
single phase qubit experiments.

We first take the decoherence times $T_1\approx658$ ns and $T_2\approx812$ ns of
each qubit in experiments of phase qubits \cite{Whittaker} as a
typical example. From Fig. 5, we know that the linear response
condition satisfies when the ramp time $t_{\text{ramp}}\gtrsim
10$ ns. So we numerically calculate the Berry curvature with
$t_{\text{ramp}}= 10$ ns and the result is plotted as the green
triangles in Fig. 4(a). In this case, the two plateaus in the Berry
curvature $F_{\theta\phi}$ are respectively near $0$ and $1$ (the
difference is about $0.96$ and the transition point remains), as expected. However, we find that the
two plateaus in the Berry curvature are gradually shifted from
$F_{\theta\phi}=0$ and $1$ when the ramp time becomes longer. For
instance, the difference between the two plateaus decreases to
about $0.72$ for the ramp time $t_{\text{ramp}}=100$ ns, which is shown as yellow squares in Fig. 4(a). This is due to the fact that the total
evolution time (i.e., 110 ns) of the system is now comparable with
the effective decoherence times (the effective times $\widetilde{T}_1\approx
658/5 $ ns and $\widetilde{T}_2 \approx 812/5$ ns) and then the Chern number
is no longer a well-defined topological index \cite{note}.
Therefore, the observation of the dynamical QHE is crucially
dependent on the long decoherence time since the Berry curvature (which associates with the Berry phase factor) has no classical
correspondence. To demonstrate the topological features of the dynamical
QHE more clearly (or in a longer ramp time), we should make improvements in coherence time for superconducting qubits in experiments. In
current technology, the relaxation time $T_1$ of the phase qubit
can be as long as $1.5$ $\mu$s \cite{Whittaker,Patel}. One can use
dynamical decoupling to increase the dephasing time up to the
$T_2=2T_1$ limit \cite{DD}. Thus, we also numerically
calculate the result for $T_1=T_2/2=1.5$ $\mu$s, % $t_{\text{ramp}}=100$
%ns and $t_{\text{meas}}=10$ ns,
and the result is plotted as black circles in Fig. 4(a). It clearly shows that the decoherence
effects are almost negligible in this case.

In conclusion, we have proposed an experimental scheme to simulate
the dynamical QHE and the related interaction-induced topological
transition with a superconducting-qubit array. We find that the
typical topological features can even be observed in the simplest
two-qubit array under practical experimental conditions.

\section{Acknowledgements}

We thank Z.-Y. Xue and C.-J. Shan for helpful discussions. This work was supported by the NSFC (Grants No. 11125417 and No.
11474153), the SKPBR of China (Grants No. 2011CB922104), and the
PCSIRT (Grant No. IRT1243). D.W.Z. acknowledges support from the postdoctoral fellowship of HKU.

\begin{widetext}
\vspace{0.5cm} ~~~~~~~~~~~~~~~~~~~~~~~~~~~\textbf{Appendix: The master equation for the two-qubit case}

\vspace{0.3cm}

In this Appendix, we derive the master equation for two-qubit array with the Hamiltonian
\begin{equation}
\nonumber
H =  - \sum\limits_{j = 1}^2 (h_j^x\sigma_x + h_j^y\sigma_y + h_j^z\sigma_z )+ J^x\sigma_1^x\sigma_2^x+ J^y\sigma_1^y\sigma_2^y+ J^z\sigma_1^z\sigma_2^z~, \eqno{\text{(A1)}}
\end{equation}
where the components of the effective magnetic field $h_1^{x,y,z} = h_2^{x,y,z} = h^{x,y,z}(t)$ are given by Eq. (3) in the text.
%\begin{equation}
%\begin{split}
%h_1^x = h_2^x = h^x\left( t \right) &= {h}\sin \theta  \cos  \phi  ,\\
%h_1^y = h_2^y = h^y\left( t \right) &= {h}\sin  \theta \sin  \phi  ,\\
%h_1^z = h_2^z = h^z\left( t \right) &= {h}\cos \theta.
%\end{split}
%\end{equation}
In the two-qubit basis $\{ |\uparrow_1\uparrow_2\rangle,~|\downarrow_1\uparrow_2\rangle,~|\uparrow_1\downarrow_2\rangle,~|\downarrow_1\downarrow_2\rangle \}$,
the Hamiltonian matrix can be written as
\begin{equation}
\nonumber
H = \left(
      \begin{array}{cccc}
        H_{11} & H_{12} & H_{13} & H_{14} \\
        H_{21} & H_{22} & H_{23} & H_{24} \\
        H_{31} & H_{32} & H_{33} & H_{34} \\
        H_{41} & H_{42} & H_{43} & H_{44} \\
      \end{array}
    \right),    \eqno{\text{(A2)}}
\end{equation}
where the matrix elements are given by
\begin{equation}
\nonumber
\begin{array}{llll}
H_{11}=-2h^z+J^z,\\ H_{22}=H_{33}=-J^z,\\ H_{44}=2h^z+J^z,\\ H_{12}=~H_{13}=~H_{24}=H_{34}=-h^x+ih^y,\\ H_{21}=~H_{31}=~H_{42}=H_{43}=-h^x-ih^y, \\ H_{23}=H_{32}=J^x+J^y, \\ H_{14}=H_{41}=J^x-J^y.
\end{array} \eqno{\text{(A3)}}
\end{equation}
The quantum dynamics of the system is described by the master equation
\begin{equation}
\nonumber
\frac{d\rho}{dt}=-i[H,\rho]+\mathcal{L}_1[\rho]+\mathcal{L}_2[\rho],
\eqno{\text{(A4)}}
\end{equation}
where the density matrix $\rho$ is denoted by
\begin{equation}
\nonumber
\rho = \left(
         \begin{array}{cccc}
           \rho_{11} & \rho_{12} & \rho_{13} & \rho_{14} \\
           \rho_{21} & \rho_{22} & \rho_{23} & \rho_{24} \\
           \rho_{31} & \rho_{32} & \rho_{33} & \rho_{34} \\
           \rho_{41} & \rho_{42} & \rho_{43} & \rho_{44} \\
         \end{array}
       \right). \eqno{\text{(A5)}}
\end{equation}
We consider the system in the Markovian and low temperature limit, and thus the Lindblad superoperator can be
written as
\begin{equation}
\nonumber
\mathcal{L}_j[\rho]=\gamma(2\sigma_j^{-}\rho\sigma_j^{+}-\{\sigma_j^{+}\sigma_j^{-},\rho\})+\Gamma(2\sigma_j^{z}\rho\sigma_j^{z}-\{\sigma_j^{z}\sigma_j^{z},\rho\})~~(j=1,2).
\eqno{\text{(A6)}}
\end{equation}
Here the relaxation rate $\gamma$ and pure dephsing rate $\Gamma$ are determined by the measured decoherence times: $1/T_1=\gamma$ and $1/T_2=1/2T_1+\Gamma$. Using the expansions
$\sigma_1^{\pm,z}\rightarrow(\sigma^{\pm,z}\otimes \mathbf{I}_{2\times2})$ and $\sigma_2^{\pm,z}\rightarrow(\mathbf{I}_{2\times2}\otimes\sigma^{\pm,z})$, one can obtain the Lindblad superoperators:
\begin{equation}
\nonumber
\begin{array}{ll}
%\mathcal{L}_1[\rho]=\left(
%                      \begin{array}{cccc}
%                       -2\gamma\rho_{11} & -2(\gamma+\Gamma)\rho_{12} & -(\gamma+2\Gamma)\rho_{13} & -(\gamma+2\Gamma)\rho_{14} \\
%                       -2(\gamma+\Gamma)\rho_{21} & -2\gamma\rho_{22} & -(\gamma+2\Gamma)\rho_{23} & -(\gamma+2\Gamma)\rho_{24} \\
%                       -(\gamma+2\Gamma)\rho_{31} & -(\gamma+2\Gamma)\rho_{32} & 2\gamma\rho_{11} & 2\gamma\rho_{12}-2\Gamma\rho_{34} \\
%                       -(\gamma+2\Gamma)\rho_{41} & -(\gamma+2\Gamma)\rho_{42} & 2\gamma_{21}-2\Gamma\rho_{43} & 2\gamma\rho_{22} \\
%                      \end{array}
%                    \right),\\\\
%\mathcal{L}_2[\rho]=\left(
%                      \begin{array}{cccc}
%                       -2\gamma\rho_{11} & -(\gamma+2\Gamma)\rho_{12} & -2(\gamma+\Gamma)\rho_{13} & -(\gamma+2\Gamma)\rho_{14} \\
%                       -(\gamma+2\Gamma)\rho_{21} & 2\gamma\rho_{11} & -(\gamma+2\Gamma)\rho_{23} & 2\gamma\rho_{13}-2\Gamma\rho_{24} \\
%                       -2(\gamma+\Gamma)\rho_{31} & -(\gamma+2\Gamma)\rho_{32} & -2\gamma\rho_{33} & -(\gamma+2\Gamma)\rho_{34} \\
%                       -(\gamma+2\Gamma)\rho_{41} & 2\gamma\rho_{31}-2\Gamma\rho_{42} & -(\gamma+2\Gamma)\rho_{43} & 2\gamma\rho_{33} \\
%                      \end{array}
%                    \right),\\\\
\mathcal{L}_1[\rho]+\mathcal{L}_2[\rho]=\left(
                      \begin{array}{cccc}
                       -4\gamma\rho_{11} & -(3\gamma+4\Gamma)\rho_{12} & -(3\gamma+4\Gamma)\rho_{13} & -(2\gamma+8\Gamma)\rho_{14} \\
                      -(3\gamma+4\Gamma)\rho_{21} & -2\gamma(\rho_{22}-\rho_{11}) & -(2\gamma+8\Gamma)\rho_{23} & 2\gamma\rho_{13}-(\gamma+4\Gamma)\rho_{24} \\
                      -(3\gamma+4\Gamma)\rho_{31} & -(2\gamma+8\Gamma)\rho_{32} & -2\gamma(\rho_{33}-\rho_{11}) & 2\gamma\rho_{12}-(\gamma+4\Gamma)\rho_{34} \\
                      -(2\gamma+8\Gamma)\rho_{41} & 2\gamma\rho_{31}-(\gamma+4\Gamma)\rho_{42} & 2\gamma\rho_{21}-(\gamma+4\Gamma)\rho_{43} & 2\gamma(\rho_{22}+\rho_{33}) \\
                      \end{array}
                    \right).

\end{array} \eqno{\text{(A7)}}
\end{equation}
By substituting Eq. (A7) into Eq. (A4), one can obtain the master equation as
\begin{equation}
\nonumber
\begin{array}{ll}
\dot{\rho}_{11}=-i\left[H_{14}(\rho_{41}-\rho_{14})+H_{12}\rho_{21}+H_{13}\rho_{31}-H_{21}\rho_{12}-H_{31}\rho_{13}\right]-4\gamma\rho_{11},\\\\

\dot{\rho}_{12}=-i\left[(H_{11}-H_{22})\rho_{12}+H_{12}(\rho_{22}-\rho_{11})+H_{13}\rho_{32}+H_{14}\rho_{42}-H_{32}\rho_{13}-H_{42}\rho_{14}\right]
-(3\gamma+4\Gamma)\rho_{12},\\\\

\dot{\rho}_{13}=-i\left[(H_{11}-H_{33})\rho_{13}+H_{13} (\rho_{33}-\rho_{11})+H_{12}\rho_{23}+H_{14}\rho_{43}-H_{23}\rho_{12}-H_{43}\rho_{14}\right]-(3\gamma+4\Gamma)\rho_{13},\\\\

\dot{\rho}_{14}=-i\left[(H_{11}-H_{44})\rho_{14}+H_{14} (\rho_{44}-\rho_{11})+H_{12}\rho_{24}+H_{13}\rho_{34}-H_{24}\rho_{12}-H_{34}\rho_{13}\right]-(2\gamma+8\Gamma)\rho_{14},\\\\

\dot{\rho}_{21}=-i\left[(H_{22}-H_{11})\rho_{21}+H_{21}(\rho_{11}-\rho_{22})+H_{23}\rho_{31}+H_{24}\rho_{41}-H_{31}\rho_{23}-H_{41}\rho_{24}\right]
-(3\gamma+4\Gamma)\rho_{21},\\\\

\dot{\rho}_{22}=-i\left[H_{23}(\rho_{32}-\rho_{23})+H_{21}\rho_{12}+H_{24}\rho_{42}-H_{12}\rho_{21}-H_{42}\rho_{24}\right]
-2\gamma(\rho_{22}-\rho_{11}),\\\\

\dot{\rho}_{23}=-i\left[(H_{22}-H_{33})\rho_{23}+H_{23} (\rho_{33}-\rho_{22})+H_{21}\rho_{13}+H_{24}\rho_{43}-H_{13}\rho_{21}-H_{43}\rho_{24}\right]-(2\gamma+8\Gamma)\rho_{23},\\\\

\dot{\rho}_{24}=-i\left[(H_{22}-H_{44})\rho_{24}+H_{24} (\rho_{44}-\rho_{22})+H_{21}\rho_{14}+H_{23}\rho_{34}-H_{14}\rho_{21}-H_{34}\rho_{23}\right]+2\gamma\rho_{13}-(\gamma+4\Gamma)\rho_{24},\\\\

\dot{\rho}_{31}=-i\left[(H_{33}-H_{11})\rho_{31}+H_{31} (\rho_{11}-\rho_{33})+H_{32}\rho_{21}+H_{34}\rho_{41}-H_{21}\rho_{32}-H_{41}\rho_{34}\right]-(3\gamma+4\Gamma)\rho_{31},\\\\

\dot{\rho}_{32}=-i\left[(H_{33}-H_{22})\rho_{32}+H_{32} (\rho_{22}-\rho_{33})+H_{31}\rho_{12}+H_{34}\rho_{42}-H_{12}\rho_{31}-H_{42}\rho_{34}\right]-(2\gamma+8\Gamma)\rho_{32},\\\\

\dot{\rho}_{33}=-i\left[H_{23}(\rho_{23}-\rho_{32})+H_{31}\rho_{13}+H_{34}\rho_{43}-H_{13}\rho_{31}-H_{43}\rho_{34}\right]
-2\gamma(\rho_{33}-\rho_{11}),\\\\

\dot{\rho}_{34}=-i\left[(H_{33}-H_{44})\rho_{34}+H_{34} (\rho_{44}-\rho_{33})+H_{31}\rho_{14}+H_{32}\rho_{24}-H_{14}\rho_{31}-H_{24}\rho_{32}\right]+2\gamma\rho_{12}-(\gamma+4\Gamma)\rho_{34},\\\\

\dot{\rho}_{41}=-i\left[(H_{44}-H_{11})\rho_{41}+H_{41} (\rho_{11}-\rho_{44})+H_{42}\rho_{21}+H_{43}\rho_{31}-H_{21}\rho_{42}-H_{31}\rho_{43}\right]-(2\gamma+8\Gamma)\rho_{41},\\\\

\dot{\rho}_{42}=-i\left[(H_{44}-H_{22})\rho_{42}+H_{42} (\rho_{22}-\rho_{44})+H_{41}\rho_{12}+H_{43}\rho_{32}-H_{12}\rho_{41}-H_{32}\rho_{43}\right]+2\gamma\rho_{31}-(\gamma+4\Gamma)\rho_{42},\\\\

\dot{\rho}_{43}=-i\left[(H_{44}-H_{33})\rho_{43}+H_{43} (\rho_{33}-\rho_{44})+H_{41}\rho_{13}+H_{42}\rho_{23}-H_{13}\rho_{41}-H_{23}\rho_{42}\right]+2\gamma\rho_{21}-(\gamma+4\Gamma)\rho_{43},\\\\

\dot{\rho}_{44}=-i\left[H_{14}(\rho_{14}-\rho_{41})+H_{42}\rho_{24}+H_{43}\rho_{34}-H_{24}\rho_{42}-H_{34}\rho_{43}\right]+2\gamma(\rho_{22}+\rho_{33}).

\end{array} \eqno{\text{(A8)}}
\end{equation}
After the evolution of the system with the ramp time $t_{\text{ramp}}$ and the total measurement time $t_{\text{meas}}$, one can obtain the final polarization along the $y$ direction at $t_f=t_{\text{ramp}}+t_{\text{meas}}$ by tracing the final density matrix governed by Eqs. (A8) as
\begin{equation}
\nonumber
\langle \sigma_1^y \rangle+\langle \sigma_2^y \rangle = \text{Tr}[\rho(t=t_f)\cdot (\sigma_y\otimes \mathbf{I}_{2\times2})]+\text{Tr}[\rho(t=t_{f})\cdot (\mathbf{I}_{2\times2}\otimes\sigma_y)]. \eqno{\text{(A9)}}
\end{equation}
For simplicity in our simulations, we treat the evolution of the qubits in the whole measurement progress as free evolution under decoherence.

\end{widetext}

\end{document}